\documentclass[aps,onecolumn,10pt,aps,amsmath,amssymb,nofootinbib,superscriptaddress,showkeys,showpacs]{revtex4-2}
\usepackage{epsfig} 
\usepackage{amsmath} 
\usepackage{graphicx}
\usepackage{color}
\usepackage{float}
\usepackage{soul,xcolor}
\usepackage[font=scriptsize]{caption}
\usepackage{braket}
\usepackage{bbold}
\usepackage{subfig}
\usepackage{braket}
\usepackage{titlesec}
\usepackage{placeins}
\usepackage{caption}
\usepackage{comment}
\usepackage[title]{appendix}%
\begin{document} 
	
	\title{Spread Complexity of High Energy Neutrino Propagation over Astrophysical Distances}

\author{Khushboo Dixit}
\affiliation{Centre for Astro-Particle Physics (CAPP) and Department of Physics, University of Johannesburg, PO Box 524, Auckland Park 2006, South Africa}
\author{S. Shajidul Haque}
\affiliation{High Energy Physics, Cosmology \& Astrophysics Theory Group and The Laboratory for Quantum Gravity \& Strings, Department of Mathematics \& Applied Mathematics, University of Cape Town, Cape Town, South Africa}
\affiliation{Department of Mathematics and Natural Sciences, Brac University, Dhaka, Bangladesh}
\affiliation{National Institute for Theoretical and Computational Sciences (NITheCS), Private Bag X1, Matieland, South Africa}
\author{Soebur Razzaque}
\affiliation{Centre for Astro-Particle Physics (CAPP) and Department of Physics, University of Johannesburg, PO Box 524, Auckland Park 2006, South Africa}
\affiliation{Department of Physics, The George Washington University, Washington, DC 20052, USA}
\affiliation{National Institute for Theoretical and Computational Sciences (NITheCS), Private Bag X1, Matieland, South Africa}

	\date{\today}
	
	\begin{abstract}
		Spread complexity measures the minimized spread of quantum states over all choices of basis. It generalizes Krylov operator complexity to quantum states under continuous Hamiltonian evolution. In this paper, we study spread complexity in the context of high-energy astrophysical neutrinos and propose a new flavor ratio based on complexity. Our findings indicate that our proposal might favor an initial ratio of fluxes as $\phi_{\nu_e}^0: \phi_{\nu_\mu}^0: \phi_{\nu_\tau}^0 = 1:0:0$ over a more generally expected ratio of $1:2:0$, when the IceCube neutrino observatory achieves its projected sensitivity to discriminate between flavors. Additionally, complexity-based definitions of flavor ratios exhibit a slight but nonzero sensitivity to the neutrino mass ordering, which traditional flavor ratios cannot capture.
	\end{abstract}
	
	\maketitle

	\section{Introduction}
	Neutrinos have long been a subject of profound interest within the field of particle physics. The discovery of neutrino oscillations, which signifies the massive nature of neutrinos, has significantly impacted the scientific community, prompting an intensified quest for physics beyond the Standard Model (BSM) \cite{Kamiokande-II:1991pyu, Cleveland:1998nv, Ahmad:2002jz, Fukuda:1994mc, Fukuda:1998mi, Abe:2011fz, DayaBay:2018yms, RENO:2018dro}. Presently, numerous long-baseline experiments, both established and planned, are dedicated to scrutinizing oscillation parameters with heightened precision and resolution \cite{DUNE:2020jqi, Zhang:2021adu, NOvA:2021nfi, T2K:2023smv}. Additionally, some experiments such as the IceCube neutrino observatory, situated at the South Pole, are actively engaged in the pursuit of high-energy astrophysical neutrinos. This pursuit is motivated by the potential for these neutrinos and their flavor composition to unveil crucial insights into their production mechanisms within celestial entities \cite{Gaisser:1994yf, Athar:2000yw, Barenboim:2003jm, Beacom:2003nh, Xing:2006uk, Lipari:2007su, Pakvasa:2007dc, Esmaili:2009dz, Lai:2009ke, Choubey:2009jq, Mena:2014sja, Palladino:2015zua, IceCube:2015rro, Bustamante:2015waa, IceCube:2017der, Song:2020nfh, IceCube:2020fpi, Ackermann:2022rqc}.

In addition, various analyses have demonstrated that neutrinos exhibit intriguing properties such as entanglement and several other nonlocal correlations \cite{Blasone:2007wp, Blasone:2007vw, Gangopadhyay:2013aha, Alok:2014gya, Banerjee:2015mha, Formaggio:2016cuh, Fu:2017hky, Song:2018bma, Ettefaghi:2020otb, Shafaq:2020sqo}. These characteristics underscore their significant potential for executing diverse tasks pertinent to Quantum Information Processing (QIP). 
One of the important aspects related to quantum information and computation theory is known as quantum complexity, which was proposed as a useful probe \cite{Susskind:2014moa} for the interior of a black hole by using the anti-de Sitter/conformal field theory (AdS/CFT) duality. Holographic complexity, proposed to be equivalent to the "volume" or ``action,'' elucidates the physics behind the horizon of an eternal AdS black hole \cite{Susskind:2014moa, Susskind:2014rva, Stanford:2014jda, Brown:2015bva, Brown:2015lvg}. However, in recent days, complexity has proven to be extremely useful for understanding various aspects of quantum systems such as quantum chaos \cite{Ali:2019zcj, Bhattacharyya:2019txx, Bhattacharyya:2020iic, Balasubramanian:2019wgd, Bhattacharyya:2020art, Balasubramanian:2021mxo}, quantum phase transitions \cite{Ali:2018aon}, quantum decoherence \cite{Bhattacharyya:2022rhm, Bhattacharyya:2021fii}, and more. It has also been applied to investigate cosmological perturbation models and the evolution of the universe \cite{Bhattacharyya:2020rpy, Bhattacharyya:2020kgu, Haque:2021hyw}.

Complexity can be described in two ways when dealing with quantum mechanical systems. First, according to Nielsen's definition, complexity is the minimal distance in the space of the unitaries between the identity and the time evolution operator $U(t)$ of the system \cite{Nielsen:2006cea, Nielsen:2005mkt, Dowling:2006tnk, Jefferson:2017sdb, Ali:2018fcz}. The second approach defines complexity as minimizing the spread of the state over a specified basis during the time evolution \cite{Balasubramanian:2022tpr, Caputa:2022eye}. More details on constructing a spread complexity measure motivated by operator complexity have been discussed in Refs.~\cite{Parker:2018yvk,Rabinovici:2020ryf,Caputa:2021sib,Muck:2022xfc,Balasubramanian:2022dnj,Bhattacharjee:2022qjw,Alishahiha:2022anw,Erdmenger:2023wjg,Hashimoto:2023swv,Camargo:2023eev,Caputa:2023vyr,Balasubramanian:2023kwd,Nandy:2024htc}.

Recently we have studied the complexity in the context of neutrino oscillations in accelerator-based long baseline neutrino experiments \cite{Dixit:2023fke}. This study not only indicated that complexity could provide more information on the CP-violating phase compared to the oscillation probability, but it also provided a perspective on the preferred value of the CP-violating phase angle. In that study we found that complexity is maximized for a $CP$-violating phase that is favored by ongoing T2K and NOvA experiments \cite{Dixit:2023fke}. The maximization of complexity has also been observed for the model based on de-sitter space which is the best suitable model to explain the inflation era \cite{Bhattacharyya:2020kgu}. Motivated by this, in this paper, we examine the complexity of astrophysical neutrinos by considering three different initial flavor ratios based on their possible production mechanisms. In this work, we have analyzed the spread complexity for astrophysical neutrinos for the first time and provided the modified definitions of flavor ratios in terms of spread complexities for each flavor.
IceCube Neutrino Observatory, currently operating in the Antarctica has been collecting data on astrophysical neutrinos and has potential to measure their flavor ratios \cite{IceCube:2017der}, especially after the recent detection of astrophysical tau neutrinos \cite{IceCube:2020fpi}. 
Our findings suggest that the modified flavor ratios, in terms of complexities, support the idea that neutron-decay (with an initial flux ratio of $\phi_{\nu_e}^0: \phi_{\nu_\mu}^0: \phi_{\nu_\tau}^0 = 1:0:0$) might be favored over $\pi$-decay (with an initial flux ratio of $1:2:0$) as the mechanism responsible for the production of neutrinos in celestial bodies. We delve into the details of neutrino oscillations and flavor ratios in the context of astrophysical neutrinos in sec. \ref{Astrophysical neutrino oscillations}, followed by defining spread complexity in sec. \ref{complexity}. We then provide the analytical expressions of flavor ratios modified in terms of complexities in sec. \ref{complexity for Astroneutrinos}. Finally, we present our results and discussions in sec. \ref{results} and \ref{discussion}, respectively.

	\section{Neutrino Oscillations and Quantum Complexity}
	In this section, we discuss neutrino oscillations in the context of astrophysical neutrinos and recall the complexity embedded in a state evolution. This is followed by a discussion on the complexities for each initial neutrino flavor modified in the context of astrophysical neutrinos. Additionally, we propose the redefined flavor ratios of neutrinos in terms of complexities.  

\subsection{Astrophysical neutrino oscillations and flavor ratios}\label{Astrophysical neutrino oscillations}
It is now well known that neutrinos can change their flavor while traveling to a distant place. Depending on their energy as well as their sensitivity to the mass squared difference associated to their mass eigenstates one can calculate the oscillation length for such flavor transitions. Neutrino oscillations occur as the flavor states of neutrino are not the mass eigenstates, but they mix via a unitary matrix to form these mass eigenstates. In the case of three flavor oscillations, the form of this unitary matrix can be given as
\begin{equation*}
    U = \begin{pmatrix}
          1      & 0      & 0\\
          0  & c_{23}  & s_{23}\\
          0  & -s_{23} & c_{23}\\
    \end{pmatrix}\begin{pmatrix}
        c_{13}  & 0  & s_{13}e^{-i\delta}\\
            0   & 1  & 0\\
        -s_{13}e^{i\delta}  & 0  & c_{13}\\
    \end{pmatrix}\begin{pmatrix}
        c_{12}  & s_{12}  & 0\\
        -s_{12}  & c_{12}  & 0\\
        0 & 0  & 1\\
    \end{pmatrix},
\end{equation*}
where, $c_{ij}$ and $s_{ij}$ are $\cos \theta_{ij}$ and $\sin \theta_{ij}$ with mixing angles $\theta_{ij}$ ($i,j=1,2,3$). The probability of survival or oscillation after the time evolution of the neutrino with a specific flavor can be given by 
\begin{align*}
    P_{\alpha\beta} = \delta_{\alpha\beta} &- 4 \sum_{j>k} Re\{U_{\alpha j}^\ast U_{\beta j} U_{\alpha k} U_{\beta k}^\ast \} \sin^2 \left(\frac{\Delta m^2_{ij}L}{4E}\right)\\
    &- 2 \sum_{j>k} Im\{U_{\alpha j}^\ast U_{\beta j} U_{\alpha k} U_{\beta k}^\ast \} \sin \left(\frac{\Delta m^2_{ij}L}{2E}\right)
\end{align*}
In the case of astrophysical neutrinos where neutrinos travel through the distances of Mpc or larger order, the sine and cosine terms carrying the phase factor get averaged out and the probabilities take the following form
\begin{equation*}
    P_{\alpha\beta} = \delta_{\alpha\beta} - 2 \sum_{j>k} Re\{U_{\alpha j}^\ast U_{\beta j} U_{\alpha k} U_{\beta k}^\ast \} 
\end{equation*} 
The quantity of interest in the experiments searching for high energy astrophysical neutrinos is the so-called {\it flavor ratio}. The production of neutrinos in astrophysical objects has been a crucial area of study for understanding the astrophysics of their sources and for the exploration of new physics. A key focus has been to determine the type of neutrinos produced at their sources. When high-energy astrophysical neutrinos reach the Earth, they undergo flavor oscillations, causing changes in the fractions of neutrinos detected on Earth based on the initial flavor ratio, see, e.g., Refs.~\cite{Athar:2000yw, Beacom:2003nh, Ackermann:2022rqc}. This suggests the existence of distinct mechanisms of neutrino production, environmental influences, and new interactions. 
The initial $\nu_{\tau}$ flavor is supposed to be zero as the neutrino production mechanism in these high energy astrophysical objects like blazars or AGNs are considered to be either neutron beta decay, $n\to p+e^-+{\bar \nu}_e$, which induces initial flavor ratio of $\phi_e^0:\phi_{\mu}^0:\phi_{\tau}^0 = 1:0:0$, or pion decay created by interactions of shock-accelerated protons/ions with surrounding radiation (p-$\gamma$) or gas (p-p), $\pi^+\to e^+ + \nu_e + {\bar \nu}_\mu + \nu_\mu$, with initial flavor ratio of $1:2:0$, see, e.g., Refs.~\cite{Gaisser:1994yf, Lipari:2007su, Winter:2014pya} 
Astrophysical environment effects such as efficient synchrotron cooling of secondary muons in strong magnetic fields can prompt the initial ratio to be 0:1:0 (muon-damped) at higher energies, however, for lower energy of muons the flavor ratio of 1:1:0 is highly probable, see e.g., Refs.~\cite{Rachen:1998fd, Kashti:2005qa, Razzaque:2005bh, Kachelriess:2006ksy, Kachelriess:2007tr, Bustamante:2015waa}.  
Interactions not involving pions, such as muon pair production and decay by two photon interactions~\cite{Razzaque:2005ds} or charm production and decay \cite{Enberg:2008jm} at the source can change the initial flavor ratios.  
Moreover, some new physics possibilities, e.g., quantum gravity effects \cite{Ahluwalia:2001xc} can also affect the production mechanism of initial flavor ratios. The same is true for Dark matter annihilation or decay. However, in this work, we focus mainly on scenarios involving standard physics
The fractions of each neutrino flavor arriving on the Earth can be expressed as follows 
\begin{align}\label{fractions}
	X_e \equiv \frac{\phi_{\nu_e}}{\phi_{\nu_e}+\phi_{\nu_{\mu}}+\phi_{\nu_{\tau}}} = \frac{P_{ee}\phi_{\nu_e}^0 + P_{\mu e}\phi_{\nu_{\mu}}^0}{\phi_{\nu_e}^0 + \phi_{\nu_{\mu}}^0 + \phi_{\nu_{\tau}}^0}\nonumber\\
 X_{\mu} \equiv \frac{\phi_{\nu_{\mu}}}{\phi_{\nu_e}+\phi_{\nu_{\mu}}+\phi_{\nu_{\tau}}} = \frac{P_{e\mu}\phi_{\nu_e}^0 + P_{\mu \mu}\phi_{\nu_{\mu}}^0}{\phi_{\nu_e}^0 + \phi_{\nu_{\mu}}^0 + \phi_{\nu_{\tau}}^0}\nonumber\\
 X_{\tau} \equiv \frac{\phi_{\nu_{\tau}}}{\phi_{\nu_e}+\phi_{\nu_{\mu}}+\phi_{\nu_{\tau}}} = \frac{P_{e\tau}\phi_{\nu_e}^0 + P_{\mu \tau}\phi_{\nu_{\mu}}^0}{\phi_{\nu_e}^0 + \phi_{\nu_{\mu}}^0 + \phi_{\nu_{\tau}}^0}
\end{align}

	\subsection{Complexity of spread of states}\label{complexity}
	In this section, we recall the formulation of the complexity of the spread of states as introduced in \cite{Balasubramanian:2022tpr,Caputa:2022eye}. The evolution of a state $\ket{\psi (0)}$ is governed by the Schrodinger equation as follows
\begin{equation*}
    i\partial_t \ket{\psi (t)} = H \ket{\psi (t)},
\end{equation*}
where, $H$ is the time independent Hamiltonian of the system and the solution to this equation gives us the time evolved state 
\begin{align*}
    \ket{\psi (t)} &= e^{-i H t} \ket{\psi (0)},\\ 
                   &= \sum_{n = 0}^{\infty} \frac{(-i H t)^n}{n!} \ket{\psi (0)},\\
                   &= \sum_{n = 0}^{\infty} \frac{(-i t)^n}{n!} \ket{\psi_n},
\end{align*}
where, $\ket{\psi_n} = H^n \ket{\psi}$. We then apply the Gram-Schmidt procedure to obtain an ordered orthonormal (Krylov) basis using the set of infinite states $\{\ket{\psi_n}\}$. The number of states spanning such a basis can be equal to or less than the dynamics of the system, however, it has been seen that in case of neutrinos, the number of basis states is exactly equal to the dimension of the system \cite{Dixit:2023fke}. We expect more complex spread of a state if the dynamics involved is chaotic. To formulate the degree of complexity a cost function can be defined as 
\begin{equation}\label{cost}
    \chi = \sum_n c_n |\langle K_n|\psi (t)\rangle|^2,
\end{equation}
where, $c_n = 0, 1, 2, \dots n$ assigning the weight to the ordered Krylov states. It has been shown in \cite{Balasubramanian:2022tpr} that the above cost function achieves its minimum for the Krylov basis, hence, this minimized cost function defined in Eq. (\ref{cost}) can be treated as the complexity embedded in the evolution of the state of the system. 
	
	\subsection{Spread complexity in flavor oscillations of astrophysical neutrinos}\label{complexity for Astroneutrinos}
	Spread complexity introduced in Eq. (\ref{cost}) has been formulated for the general case of three flavor neutrino oscillations in \cite{Dixit:2023fke}. The complexity for neutrino starting as $\nu_e$ flavor at the initial point can be expressed below
\begin{align}
	\chi_e = & P_{e\mu}(t) \left[ N_{1e}^2|a_1|^2 + 2 N_{2e}^2|b_1|^2 \right] 
	+ P_{e\tau}(t) \left[ N_{1e}^2|a_2|^2 + 2 N_{2e}^2 |b_2|^2 \right] \nonumber \\
	& + 2 \Re\left[ N_{1e}^2 a_1^\ast a_2 A_{e\mu}(t) A_{e\tau}(t)^\ast \right] 
	+ 4 \Re \left[ N_{2e}^2 b_1^\ast b_2 A_{e\mu}(t) A_{e\tau}(t)^\ast \right].
	\label{eq:nue_cost}
\end{align}
Here, $P_{e\mu}$ and $P_{e\tau}$ are the oscillation probabilities for $\nu_e \rightarrow \nu_{\mu}$ and $\nu_e \rightarrow \nu_{\tau}$ transitions, respectively and $A_{e\mu}$ and $A_{e\tau}$ are the corresponding oscillation amplitudes. In the case of astrophysical neutrinos where the travel distance is significantly large (of the order of Mpc or larger), the sin and cos terms carrying the phase factor depending on the mass squared differences $m_j^2 - m_i^2$ ($j>i$ and $i,j=1,2,3$) corresponding to the mass eigenstates driving flavor oscillations get averaged out. Therefore, these probabilities and amplitudes take the following form

\begin{align*}
	P_{e\mu} =& -2\left(\Re(U_{e2}^\ast U_{\mu 2}U_{e1}U_{\mu 1}^\ast)+\Re(U_{e3}^\ast U_{\mu 3}U_{e1}U_{\mu 1}^\ast)+\Re(U_{e3}^\ast U_{\mu 3}U_{e2}U_{\mu 2}^\ast)\right)\nonumber \\
	=& \frac{1}{8} \sin \theta_{12} \cos \theta_{12} \cos ^2\theta_{13} \left(4 \cos \delta  \cos 2 \theta_{12} \sin \theta_{13} \sin 2 \theta_{23}\right.\nonumber\\
	&\left.+\sin 2 \theta_{12} \left(2 \cos 2 \theta_{13} \sin ^2\theta_{23} + 3 \cos 2 \theta_{23} + 1\right)\right),\nonumber\\
	P_{e\tau} =& -2\left(\Re(U_{e2}^\ast U_{\tau 2}U_{e1}U_{\tau 1}^\ast)+\Re(U_{e3}^\ast U_{\tau 3}U_{e1}U_{\tau 1}^\ast) + \Re(U_{e3}^\ast U_{\tau 3}U_{e2}U_{\tau 2}^\ast)\right)\nonumber\\
	=& -\cos ^2\theta_{13} \left(\frac{1}{4} \sin \theta_{13} \left(\cos \delta (4 \sin 2 \theta_{12} + \sin 4 \theta_{12}) \sin 2 \theta_{23}\right.\right.\nonumber\\ 
 &\left.\left.+ 8 \sin ^2\theta_{12} \sin \theta_{13} \cos ^2\theta_{23}\right)-2 \cos ^4\theta_{12} \sin ^2\theta_{13} \cos ^2\theta_{23}\right.\nonumber\\
 &\left.-2 \sin ^2\theta_{12} \cos ^2\theta_{12} \sin ^2\theta_{23}\right),
 \end{align*}
 \begin{align}
 A_{e\mu} A_{e\tau}^\ast =& |U_{e1}|^2 U_{\mu 1}^\ast U_{\tau 1} + |U_{e2}|^2 U_{\mu 2}^\ast U_{\tau 2} + |U_{e3}|^2 U_{\mu 3}^\ast U_{\tau 3}\nonumber\\
	=& \frac{1}{16} \cos ^2\theta_{13} (\sin 2 \theta_{23} (-(\cos 4\theta_{12} + 7) \cos 2 \theta_{13} + 3 \cos 4\theta_{12} + 5)\nonumber\\
	&+ 4 \sin 4 \theta_{12} \sin \theta_{13} (\cos \delta \cos 2 \theta_{23} + i \sin \delta)).
\end{align}
The effect of averaging out the sine and cosine terms will be visible in probabilities $P_{e\mu}$ \& $P_{e\tau}$ and the cross term $A_{e\mu} A_{e\tau}^\ast$ only. Other terms such as $|a_1|^2$, $|a_2|^2$, $|b_1|^2$, $|b_2|^2$, $a_1^\ast a_2$ and $b_1^\ast b_2$ are given in the appendix. 

One important point should also be noted that since the oscillation terms carrying the phase factor in terms of $\Delta m^2$ get averaged out, the discrimination between the normal ordering (NO) and inverted ordering (IO) of neutrino mass eigenstates is impossible through probabilities. However, complexities have slight but non-vanishing discrimination in the effects of NO and IO as can be seen in the expressions of $|a_1|^2$, $|a_2|^2$, $|b_1|^2$, $|b_2|^2$ and so on in Eq. (\ref{extra}). Similarly, the complexities associated with the evolution of neutrino states can be obtained in case of initial $\nu_{\mu}$ and $\nu_{\tau}$ flavors as well.

\begin{figure*}[ht] 
    \centering
    \begin{tabular}{cc}
        \includegraphics[width=.95\textwidth]{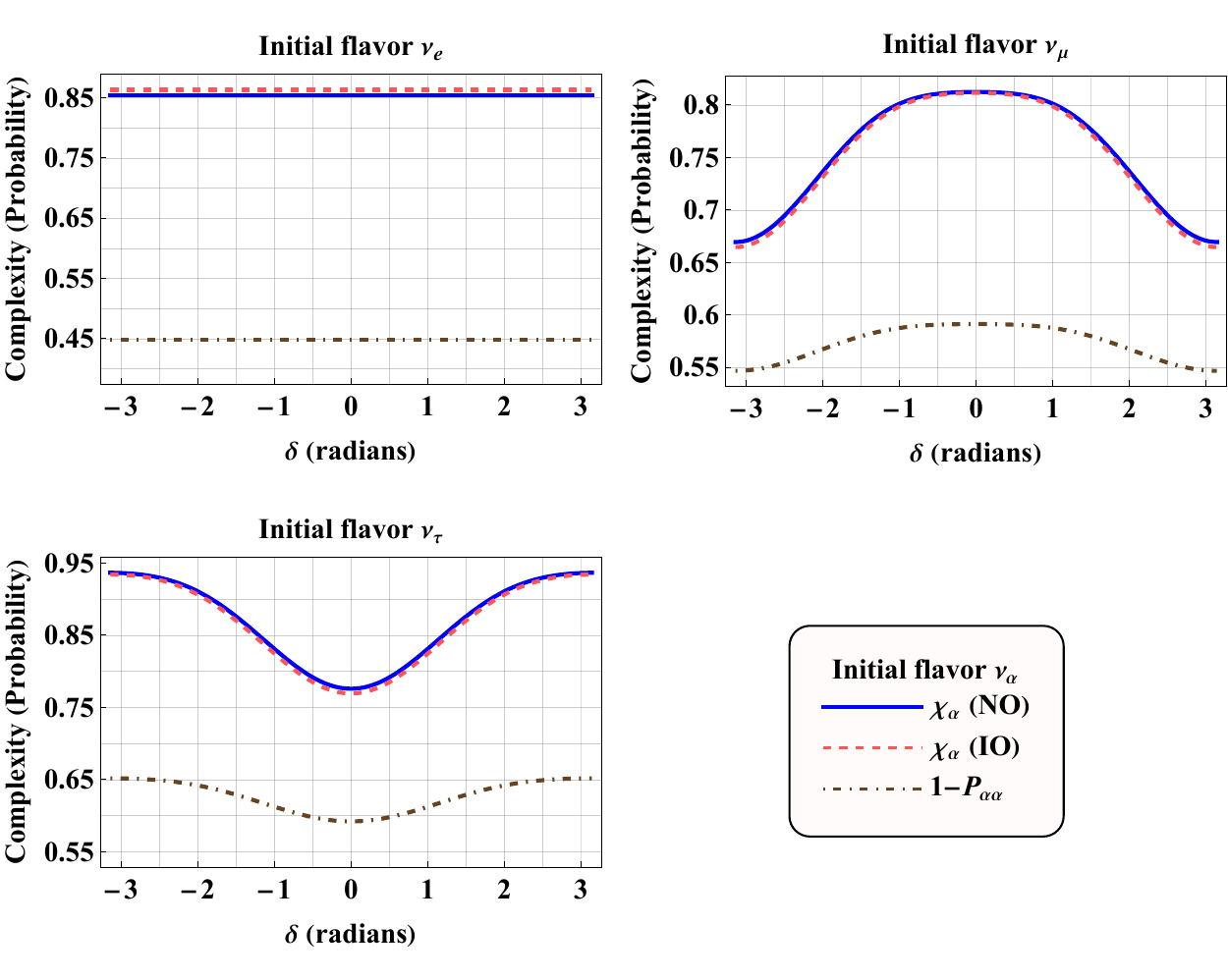} 
    \end{tabular}
    \caption{Complexities $\chi_{\alpha}$ (normal ordering (NO) and inverted ordering (IO) of neutrino masses in blue and red curves, respectively) and $1-P_{\alpha \alpha}$ (black curve) plotted with respect to the $CP$-violating phase $\delta$. Here $\alpha = e,\mu,\tau$ depends on the initial neutrino flavor. All other parameters are set to their best-fit values provided in a recent analysis \cite{Esteban:2020cvm}.}
    \label{Chi_delta}
\end{figure*}
	
	\subsection{Redefining Flavor Ratios Using Complexity}
	In this section, we will propose a new flavor ratio based on complexity. We can consider the initial flavor ratio to be $\phi_{\nu_e}^0 : \phi_{\nu_{\mu}}^0 : \phi_{\nu_{\tau}}^0 = (x : 1-x : 0) \phi_{\nu_e}^0$, where $x = 0, 1, 1/3$ which denotes flavor ratios $\phi_{\nu_e}^0 : \phi_{\nu_{\mu}}^0 : \phi_{\nu_{\tau}}^0 = (0:1:0), ~(1:0:0), ~(1:2:0)$, respectively. Here, $\phi_{\nu_e}+\phi_{\nu_{\mu}}+\phi_{\nu_{\tau}} = \phi_{\nu_e}^0+\phi_{\nu_{\mu}}^0+\phi_{\nu_{\tau}}^0$ as a consequence of probability conservation. Hence, fractions of distinct flavors defined in Eq. (\ref{fractions}) can be revised as
\begin{align}\label{modifiedfraction}
	X_e &= x P_{ee} + (1-x) P_{\mu e}\nonumber\\
 X_{\mu} &= x P_{\mu e} + (1-x) P_{\mu\mu}\nonumber\\
 X_{\tau} &= (2x-1) P_{e\tau} + (1-x) (1 - P_{\tau\tau}).
\end{align}
Then we introduce here the analogous equation to the Eq. (\ref{modifiedfraction}) in terms of complexities as
\begin{align}\label{Xprimee}
	X^\prime_e &= x (1 - \chi_e) + (1-x) P_{\mu e}.\nonumber\\
X^\prime_\mu &= x P_{\mu e} + (1-x) (1 - \chi_{\mu})\nonumber\\
 X^\prime_{\tau} &= (2x-1) P_{e\tau} + (1-x) \chi_{\tau}.
\end{align}
Here, we have replaced $1-P_{\alpha\alpha}$ with corresponding $\chi_{\alpha}$, for flavor $\alpha = e, \mu, \tau$. We take inspiration from the findings in \cite{Dixit:2023fke} and Fig.~\ref{Chi_delta} to make this replacement, noting that $\chi_\alpha$ and $1-P_{\alpha\alpha}$ share similar characteristics. The flavor ratios outlined in Eq. (\ref{Xprimee}) represent a combination of complexities and oscillation probabilities. Despite this, they can provide valuable insights, as complexities are known to encapsulate a greater amount of pertinent information about the system.
We note that the measurable quantities in the high energy neutrino telescopes are the neutrino event numbers of different flavors. These events are used to construct experimentally measured flavor ratios that can be then compared with theoretical predictions of flavor ratios based on probabilities or complexities of distinct flavor-state evolution. We compare the theoretical predictions for flavor ratios with data in the next section.

	\section{Results}\label{results}
	In this section, we initially present our results concerning flavor ratios using traditional probability-based definitions, followed by a modification of these definitions to incorporate complexities for each neutrino flavor. In \cite{Dixit:2023fke}, a significant characteristic of complexities has been identified, indicating that the complexities $\chi_{\alpha}$, $(\alpha = e, \mu, \tau)$ emulate the traits of their corresponding total oscillation probabilities, i.e., $1-P_{\alpha\alpha}$, although $\chi_{\alpha}$ provides more information. A similar observation is evident in the context of astrophysical neutrinos, as illustrated in Fig. \ref{Chi_delta}. It becomes apparent that both complexities $\chi_e$ (depicted as blue(solid) and red(dashed) curves for normal and inverted mass ordering, respectively) and the total oscillation probability $1-P_{ee}$ (denoted by the brown(dot-dashed) curve) are not impacted by the CP-phase $\delta$. However, $\chi_\mu$, $\chi_\tau$, as well as $1-P_{\mu\mu}$ and $1-P_{\tau\tau}$, exhibit variations in relation to the $\delta$ phase. Furthermore, these quantities display symmetric characteristics around $\delta = 0$, signifying their CP-even nature, where CP-even denotes $f(\delta) = f(-\delta)$, with $f$ encompassing $\chi_{\alpha}$ and $1-P_{\alpha\alpha}$. Notably, complexities $\chi_{\mu}$ and $\chi_{\tau}$ demonstrate enhanced sensitivity to the CP-phase. Additionally, Fig. \ref{Chi_delta} highlights a subtle sensitivity of complexities $\chi_{\alpha}$ to the normal and inverted mass ordering, a feature absent in the case of $1-P_{\alpha\alpha}$.

\begin{figure}[htb]
  \centering
    \includegraphics[width=0.72\textwidth, height=0.56\textwidth]{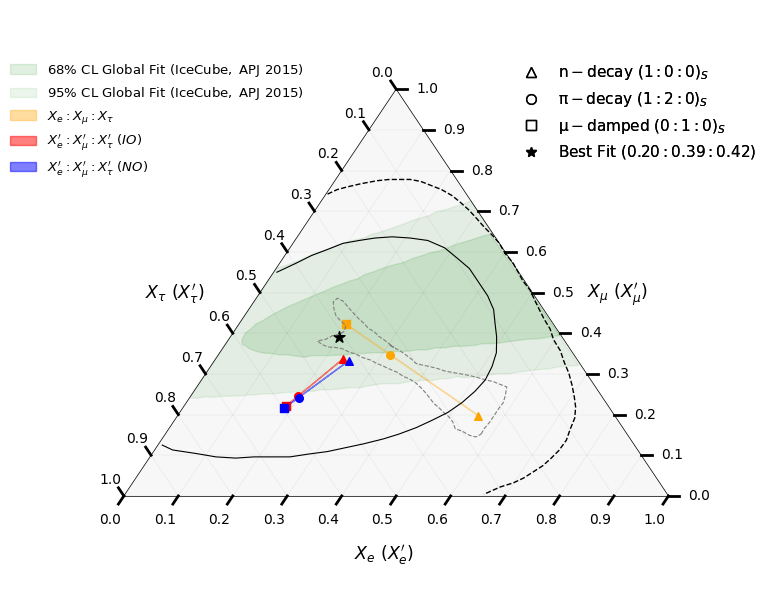}
      \caption[Images]{The flavor ratios $X_\alpha$ ($X^{\prime}_\alpha$) are shown on a ternary plot representation. The IceCube best-fit point is represented as a star mark in the plot while the contours depicting 68\% and 95\% allowed regions are shown as the solid and dashed black curves, respectively \cite{IceCube:2020fpi}. Dark-green (68\% CL) and light-green (95\% CL) regions show the global fit provided by the IceCube collaboration \cite{IceCube:2015gsk}. The gray dashed lines show the 3$\sigma$ allowed region of oscillation parameters. Blue and red colors depict the cases of relative fractions of $X^\prime_{e}$, $X^\prime_{\mu}$ and $X^\prime_{\tau}$ corresponding to the normal ordering (NO) and inverted ordering (IO), respectively. The orange points represent the fraction of traditionally defined flavor ratios, $i.e.,$ $X_e$, $X_{\mu}$ and $X_{\tau}$. Solid lines (purple, blue and red) represent the flavor ratios with varying $x$ in the range [0,1]. The cases of initial flavor ratios ($1:0:0$), ($0:1:0$) and ($1:2:0$) corresponding to $x = 1, 0, 1/3$, $i.e.,$ respectively are shown in different styles of points as mentioned in the legend.}
      \label{fig:I1}
      \end{figure}

    \begin{figure}[htb]
     \centering
    \includegraphics[width=.8\textwidth]{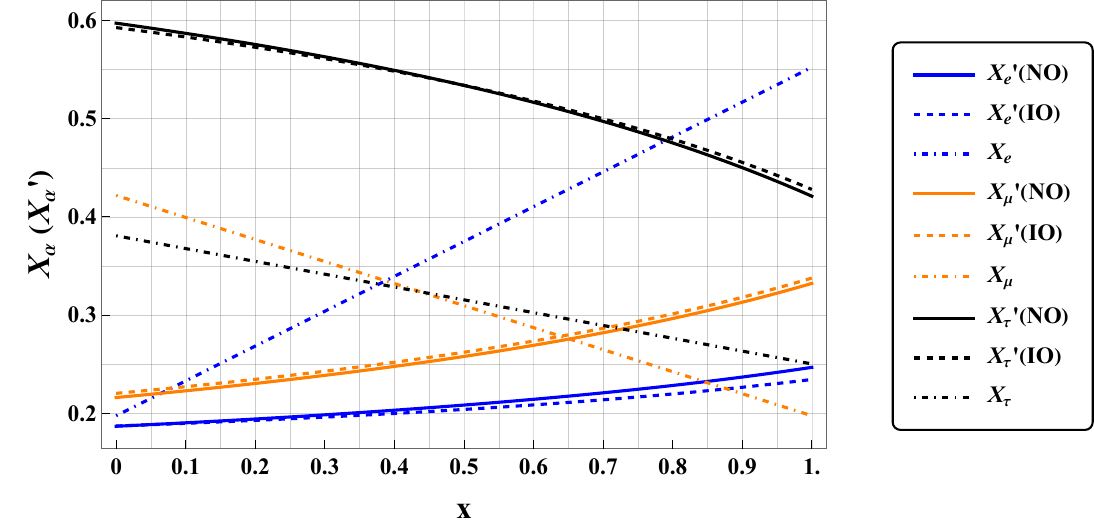}
      \caption[Image 2]{$X^{\prime}_e$, $X^{\prime}_{\mu}$ and $X^{\prime}_{\tau}$ defined in Eq. (\ref{Xprimee}) are plotted with respect to $x$. The solid and dashed curves represent $X^{\prime}_{\alpha}$ with NO and $X^{\prime}_{\alpha}$ with IO, while the associated initial neutrino flavor $\nu_e$, $\nu_\mu$ and $\nu_\tau$ are represented by blue, orange and black color, respectively. Dot-dashed curves represent traditional ratios $X_{\alpha}$ defined in Eq. (\ref{modifiedfraction}), respectively.}
  \label{fig:images}
\end{figure}

In Fig. \ref{fig:I1}, the ternary plot shows the representation of flavor ratios. The traditional flavor ratios \(X_{\alpha}\) are depicted in orange, while their analogous quantities \(X^\prime_{\alpha}\) are shown in blue (for NO) and red (for IO). The green shaded regions represent the global fit results provided by IceCube \cite{IceCube:2015gsk} and the recent best-fit results in black color \cite{IceCube:2020fpi}.

The data from the IceCube experiment so far indicates consistency with the initial flavor ratios of $1:2:0$ that appear as the ratios $1:1:1$ on Earth \cite{IceCube:2015rro, IceCube:2020fpi, IceCube:2013low}. However, the best-fit point (see Fig.~\ref{fig:I1}) gives a ratio of $0.20:0.39:0.42$ \cite{IceCube:2020fpi}, which corresponds to an initial flavor ratio close to $0:1:0$ with the traditional probability-based flavor ratio expectation. In contrast, the complexity-based expected flavor ratios favor the initial ratio of neutrino flavors to be $1:0:0$.  The findings in reference \cite{Mena:2014sja} indicate that the best fit to IceCube data at the time is achieved with a $1:0:0$ ratio at Earth, which does not align with any of these proposed initial flavor ratios, namely ($1:0:0$), ($0:1:0$), ($1:2:0$), or any other arbitrary choice as shown in Fig. \ref{fig:images} where we vary the initial fraction via the parameter $x$ between [0,1]. 

Apart from the above observations, the following points are important to be mentioned: 
\begin{itemize}
	\item Firstly, the analogous flavor ratios $X^\prime_\alpha$ defined in terms of complexities $\chi_e$, $\chi_{\mu}$ and $\chi_{\tau}$ can have more spread with respect to the variation in $CP$-phase $\delta$ both in the case of NO and IO in comparison to the traditional quantities $X_\alpha$ as the complexities in Fig. (\ref{Chi_delta}) are shown to have more variations with respect to $\delta$ than the total oscillation probabilities. That means these newly defined flavor ratios can provide better resolution for the $CP$-violating phase value. Therefore, complexity-based flavor ratios can also offer the possibility to constrain or exclude a significant range of the $\delta$-phase, in principle.
	\item Secondly, in Fig. (\ref{fig:I1}), the newly defined flavor ratios in terms of complexities also show small but non-zero sensitivity to the neutrino mass ordering (slight difference between the blue and red lines). Specifically, the initial flavor ratio $\phi_{\nu_e}^0:\phi_{\nu_\mu}^0:\phi_{\nu_\tau}^0 = 1:0:0$ favors this condition the most. This sensitivity is absent in the case of traditional flavor ratios (purple). The point to be noticed here is that complexity-based ratios also favor the case of $1:0:0$ over $1:2:0$ as initial flavor ratios of the astrophysical neutrinos, while compared with the sensitivity contours shown in the plot.
\end{itemize}
The above-mentioned advantages of the redefined flavor ratios, $X^\prime_\alpha$ in Eq. (\ref{Xprimee}) in terms of the complexities over the traditional ones, $X_\alpha$, in Eq. (\ref{modifiedfraction}) can be attributed to the cross terms present in the complexities defined in Eq. (\ref{eq:nue_cost}) along with the terms proportional to the oscillation probabilities. We also provide individual fractions of neutrino flavors varying with respect to arbitrary values of $x$ (the parameter defining the initial neutrino flavor ratio) in Fig. (\ref{fig:images}), which may correspond to non-traditional neutrino production mechanism.
	
	\section{Discussion}\label{discussion}
	In this work, we have attempted to redefine the neutrino flavor ratios arriving on the Earth in terms of the complexities of neutrino propagation due to the motivation that these complexities and the total oscillation probabilities of neutrino flavor states depict equivalent features, whereas the former convey more information. Comparing these complexity-based and traditional flavor ratios with the IceCube global fit (Fig.~\ref{fig:I1}, green shaded regions), we find that the initial flavor ratio of $1:0:0$, which might indicate the production mechanism of neutrinos to be from neutron decays, is favored over other ratios in the complexity-based model but is ruled out at 95\% CL in the traditional model.  
We also note that while the error contours of the data are still large, more robust measurement of these ratios is expected in future with large data samples collected by neutrino telescopes such as IceCube-Gen2 \cite{IceCube-Gen2:2020qha}, KM3NeT/ARCA \cite{Biagi:2023yoy}, Baikal-GVD \cite{Baikal-GVD:2023beh}, P-One \cite{Malecki:2024tvt} and Hyper-Kamiokande \cite{Abe:2024kjq}. More data from current and upcoming neutrino telescopes will shrink the error contours and provide meaningful constraints on models. Furthermore, the complexity-based flavor ratios are found to be sensitive to the neutrino mass ordering, unlike the case of traditional flavor ratios. Also, notably the complexity-based flavor ratios manifest a greater degree of variability in the $CP$-violating phase. These features might be tested in the future with observational data from the neutrino telescopes. 

The concept of spread complexity is fundamental and applicable to any quantum system for analyzing the complexity in its state's time evolution. We do not foresee further changes to this definition, but an analogous quantity for neutrino systems incorporating the flavor basis instead of Krylov could be proposed. However, this would not minimize the complexity cost function as Krylov does. Our results show that spread complexity is sensitive to mass ordering in the neutrino system, due to Krylov states being mixtures of all three flavors, making it more informative than traditional flavor probabilities. While defining flavor ratios solely based on complexity is possible and necessary, it is beyond the scope of this work.

We would like to highlight another aspect for estimating the complexity embedded in a system, specifically in terms of Krylov-entropy or K-entropy \cite{Balasubramanian:2022tpr, Barbon:2019wsy}. This measure of complexity differs from Krylov complexity, and from a preliminary study we found that K-entropy demonstrates a logarithmic dependence on oscillation probabilities. 
Nevertheless, the overall optimization characteristics are quite similar to those of Krylov complexity. Additionally, this entropic measure of complexity might reveal other hidden aspects of the system that need to be explored in more detail. We will report those findings in a future work.

	\begin{acknowledgements}
		This work was partially supported by grants from the National Research Foundation (NRF), South Africa, through the National Institute of Theoretical and Computational Sciences (NITheCS) and from the University of Johannesburg Research Council.
	\end{acknowledgements}

\begin{appendices}
\section{}\label{secA1} 
\begin{eqnarray}\label{extra}
	|a_1|^2 =& (\Delta m_{21}^2)^2 |U_{e2}|^2 |U_{\mu 2}|^2 + (\Delta m_{31}^2)^2 |U_{e3}|^2 |U_{\mu 3}|^2 + 2 (\Delta m_{21}^2)(\Delta m_{31}^2) \Re(U_{e2}^\ast U_{\mu 2} U_{e3}U_{\mu 3}^\ast)\nonumber \\
	=& \cos ^2 \theta_{13} \left((\Delta m^2_{21})^2 \sin ^2 \theta_{12} \left(\sin ^2 \delta \sin ^2 \theta_{12} \sin ^2 \theta_{13} \sin ^2\theta_{23}\right.\right.\nonumber\\
	&\left.\left. + \left(\cos \theta_{12} \cos \theta_{23} - \cos \delta \sin \theta_{12} \sin \theta_{13} \sin \theta_{23}\right)^2\right)  + \Delta m^2_{31} \sin \theta_{13} \sin \theta_{23} \right.\nonumber\\
	&\left. \left( \Delta m^2_{21} \cos \delta \sin 2 \theta_{12} \cos \theta_{23} + \sin \theta_{13} \sin \theta_{23} \left(\Delta m^2_{31} - 2 \Delta m^2_{21} \sin ^2\theta_{12}\right)\right)\right)\nonumber \\
	|a_2|^2 =& (\Delta m_{21}^2)^2 |U_{e2}|^2 |U_{\tau 2}|^2 + (\Delta m_{31}^2)^2 |U_{e3}|^2 |U_{\tau 3}|^2 + 2 (\Delta m_{21}^2)(\Delta m_{31}^2) \Re(U_{e2}^\ast U_{\tau 2} U_{e3}U_{\tau 3}^\ast)\nonumber \\
	=& \cos ^2\theta_{13} \left(-\frac{1}{4} \Delta m^2_{21} \cos \delta \sin 2 \theta_{12} \sin \theta_{13} \sin 2 \theta_{23} (\Delta m_{21}^2 \cos 2 \theta_{12} - \Delta m_{21}^2 + 2 \Delta m_{31}^2)\right.\nonumber\\
	&\left.+ (\Delta m_{21}^2)^2 \sin ^2\theta_{12} \cos ^2\theta_{12} \sin ^2\theta_{23}\right.\nonumber\\
	&\left. +\frac{1}{4} \sin ^2\theta_{13} \cos ^2\theta_{23} (\Delta m_{21}^2 (-\cos 2 \theta_{12}) + \Delta m_{21}^2 - 2 \Delta m_{31}^2)^2\right)\nonumber\\
	|b_1|^2 =& (\Delta m_{21}^2)^2 (\Delta m_{21}^2 - 2E_{\nu}A_e)^2 |U_{e2}|^2 |U_{\mu 2}|^2 + (\Delta m_{31}^2)^2 (\Delta m_{31}^2 - 2E_{\nu}A_e)^2 |U_{e3}|^2 |U_{\mu 3}|^2\nonumber\\
	& + 2 (\Delta m_{21}^2)(\Delta m_{31}^2)(\Delta m_{21}^2 - 2E_{\nu}A_e)(\Delta m_{31}^2 - 2E_{\nu}A_e) \Re(U_{e2}^\ast U_{\mu 2} U_{e3}U_{\mu 3}^\ast)\nonumber \\
	|b_2|^2 =& (\Delta m_{21}^2)^2 (\Delta m_{21}^2 - 2E_{\nu}A_e)^2 |U_{e2}|^2 |U_{\tau 2}|^2 + (\Delta m_{31}^2)^2 (\Delta m_{31}^2 - 2E_{\nu}A_e)^2 |U_{e3}|^2 |U_{\tau 3}|^2\nonumber\\
	& + 2 (\Delta m_{21}^2)(\Delta m_{31}^2)(\Delta m_{21}^2 - 2E_{\nu}A_e)(\Delta m_{31}^2 - 2E_{\nu}A_e) \Re(U_{e2}^\ast U_{\tau 2} U_{e3}U_{\tau 3}^\ast)\nonumber\\
	a_1^\ast a_2 =& (\Delta m_{21}^2)^2 |U_{e2}|^2 U_{\mu 2}^\ast U_{\tau 2} + (\Delta m_{31}^2)^2 |U_{e3}|^2 U_{\mu 3}^\ast U_{\tau 3} \nonumber\\
	&+ (\Delta m_{21}^2)(\Delta m_{31}^2) [U_{e2}U_{\mu 2}^\ast U_{e3}^\ast U_{\tau 3} + U_{e3}U_{\mu 3}^\ast U_{e2}^\ast U_{\tau 2}]\nonumber\\
	=& \frac{1}{8} \cos ^2\theta_{13} \left(-8 e^{-i \delta } \Delta m^2_{21} \sin \theta_{12} \cos \theta_{12} \sin \theta_{13} \left(-\sin ^2\theta_{23} + e^{2 i \delta} \cos ^2\theta_{23}\right)\right.\nonumber\\
	&\left.\left(\Delta m^2_{21} \sin ^2\theta_{12} - \Delta m^2_{31}\right)- (\Delta m^2_{21})^2 \sin ^2 2 \theta_{12} \sin 2 \theta_{23}\right.\nonumber\\
	&\left.+\sin ^2\theta_{13} \sin 2 \theta_{23} (2\Delta m^2_{21} \sin^2 \theta_{12} - 2 \Delta m^2_{31})^2\right),\nonumber\\
	b_1^\ast b_2 =& (\Delta m_{21}^2)^2 (\Delta m_{21}^2 - 2E_{\nu}A_e)^2 |U_{e2}|^2 U_{\mu 2}^\ast U_{\tau 2} + (\Delta m_{31}^2)^2 (\Delta m_{31}^2 - 2E_{\nu}A_e)^2 |U_{e3}|^2 U_{\mu 3}^\ast U_{\tau 3}\nonumber\\
	& + (\Delta m_{21}^2)(\Delta m_{31}^2)(\Delta m_{21}^2 - 2E_{\nu}A_e)(\Delta m_{31}^2 - 2E_{\nu}A_e) [U_{e2}U_{\mu 2}^\ast U_{e3}^\ast U_{\tau 3} + U_{e3}U_{\mu 3}^\ast U_{e2}^\ast U_{\tau 2}]\nonumber\\
	=& \left[e^{-i \delta } (\Delta m^2_{21})^2 (\Delta m^2_{31})^2 (\Delta m^2_{21} - \Delta m^2_{31})^2 \sin ^2 \theta_{12} \cos ^2 \theta_{12} \sin ^2\theta_{13} \cos ^2\theta_{13} \left(\Delta m^2_{21} \sin \theta_{12} \right.\right.\nonumber\\
	&\left.\left.\cos \theta_{12} \sin \theta_{23} - e^{i \delta } \sin \theta_{13} \cos \theta_{23} \left(\Delta m^2_{31} - \Delta m^2_{21} \sin ^2\theta_{12}\right)\right)\right.\nonumber\\
	&\left. \left(\sin \theta_{13} \sin \theta_{23} \left(\Delta m^2_{31} - \Delta m^2_{21} \sin ^2\theta_{12}\right) + e^{i \delta } \Delta m^2_{21} \sin \theta_{12} \cos \theta_{12} \cos \theta_{23}\right)\right]\nonumber\\
	&\times \left[\left(-(\Delta m^2_{21})^2 \sin ^4\theta_{12} \cos ^2\theta_{13} + \Delta m^2_{21} \sin ^2\theta_{12}\right.\right.\nonumber\\
	&\left.\left. (\Delta m^2_{21} + \Delta m^2_{31} \cos 2 \theta_{13} - \Delta m^2_{31}) + (\Delta m^2_{31})^2 \sin ^2 \theta_{13}\right)^2\right].
\end{eqnarray}
The normalization constants can be spelled as
\begin{align}
	N_{1e}^2 = &\left[(\Delta m_{21}^2)^2 |U_{e2}|^2(1-|U_{e2}|^2) + (\Delta m_{31}^2)^2 |U_{e3}|^2(1-|U_{e3}|^2)\right. \nonumber\\
	&\left.- 2(\Delta m_{21}^2)(\Delta m_{31}^2)|U_{e2}|^2|U_{e3}|^2 \right]^{-1}\\
	N_{2e}^2 =& \left[(\Delta m_{21}^2)^2 (\Delta m_{21}^2-2 E_{\nu}A_e)^2 |U_{e2}|^2(1-|U_{e2}|^2) + (\Delta m_{31}^2)^2 (\Delta m_{31}^2-2 E_{\nu}A_e)^2 |U_{e3}|^2\right.\nonumber\\
	&\left. (1-|U_{e3}|^2) - 2(\Delta m_{21}^2)(\Delta m_{31}^2)(\Delta m_{21}^2-2 E_{\nu}A_e)(\Delta m_{31}^2-2 E_{\nu}A_e)|U_{e2}|^2|U_{e3}|^2 \right]^{-1}
\end{align}
where,
\begin{align*}
	A_e = \frac{1}{2E}&\left[(\Delta m_{21}^2)^3 |U_{e2}|^2 (1-|U_{e2}|^2) + (\Delta m_{31}^2)^3 |U_{e3}|^2 (1-|U_{e3}|^2) - \Delta m_{21}^2 \Delta m_{31}^2 |U_{e2}|^2 |U_{e3}|^2\right. \nonumber\\
	&\left.(\Delta m_{21}^2+\Delta m_{31}^2)\right]
	\left[(\Delta m_{21}^2)^3 |U_{e2}|^2 (1-|U_{e2}|^2) + (\Delta m_{31}^2)^3 |U_{e3}|^2 (1-|U_{e3}|^2)\right.\nonumber\\
	&\left.-2 (\Delta m_{21}^2) (\Delta m_{31}^2)|U_{e2}|^2 |U_{e3}|^2\right]^{-1}
\end{align*} 

\end{appendices}

\end{document}